\documentclass[runningheads]{llncs}
\usepackage{graphicx}
\usepackage{multirow,amsmath}
\usepackage{color}
\usepackage{tabu, floatrow, makecell, booktabs, tabularx}
\newcommand*{\MyIndent}{\hspace*{0.3cm}}

\newcommand{\ie}{\textit{i}.\textit{e}., }
\newcommand{\eg}{\textit{e}.\textit{g}., }

\begin{document}
\title{E$^2$Net: An Edge Enhanced Network for Accurate Liver and Tumor Segmentation on CT Scans}
\titlerunning{An Edge Enhanced Network for Accurate Liver and Tumor Segmentation}
%
\author{Youbao Tang\inst{1} \and
Yuxing Tang\inst{1} \and
Yingying Zhu\inst{1} \and
Jing Xiao\inst{2} \and \\
Ronald M. Summers\inst{1}}
%
\authorrunning{Youbao Tang et al.}
%
\institute{Imaging Biomarkers and Computer-Aided Diagnosis Laboratory, Radiology and Imaging Sciences, National Institutes of Health Clinical Center, Bethesda, MD 20892-1182, USA \\ \email{youbao.tang@nih.gov, rms@nih.gov} \and
Ping An Insurance Company of China, Shenzhen, 510852, China}

%
%
\maketitle 
\begin{abstract}
Developing an effective liver and liver tumor segmentation model from CT scans is very important for the success of liver cancer diagnosis, surgical planning and cancer treatment.
In this work, we propose a two-stage framework for 2D liver and tumor segmentation. The first stage is a coarse liver segmentation network, while the second stage is an edge enhanced network (E$^2$Net) for more accurate liver and tumor segmentation. E$^2$Net explicitly models complementary objects (liver and tumor) and their edge information within the network to preserve the organ and lesion boundaries. 
We introduce an edge prediction module in E$^2$Net and design an edge distance map between liver and tumor boundaries, which is used as an extra supervision signal to train the edge enhanced network. We also propose a deep cross feature fusion module to refine multi-scale features from both objects and their edges. 
E$^2$Net is more easily and efficiently trained with a small labeled dataset, and it can be trained/tested on the original 2D CT slices (resolve resampling error issue in 3D models). The proposed framework has shown superior performance on both liver and liver tumor segmentation compared to several state-of-the-art 2D, 3D and 2D/3D hybrid frameworks.

\end{abstract}
\setcounter{footnote}{0}
\section{Introduction}

The liver is the body's largest internal organ and liver cancer is the leading cause of cancer deaths worldwide, accounting for more than 700,000 deaths each year\footnote{\url{https://www.cancer.org/cancer/liver-cancer.html}}.
Computed tomography (CT) is commonly adopted for imaging abdominal organs including liver.
Developing accurate, robust and automated techniques for liver and its tumor segmentation from CT scans is of high demand to assist clinicians in liver cancer diagnosis, surgical planning and precision medicine in clinical practice.
However, liver and liver tumor segmentation is very challenging due to low contrast or blurry/unclear boundaries between the liver, tumor and nearby organ tissues.
Moreover, the pathology of liver tumor is inherently heterogeneous on population, which leads to large variations on the size, shape, location, appearance/textures and numbers of tumors within one patient. 
In the last decade, deep learning has been successfully and widely used in many tasks of medical image analysis \cite{cai2018accurate,tang2020automated,tang2019uldor,agarwal2020weakly,tang2019tuna,jin2018ct,tang2018semi,pelvic_yirui,tang2019ct,yan2019mulan,tang2019xlsor,zhu2020cross,tang2018ct,tang2019abnormal,chen2020anatomy,tang2020one}. Also, researchers have developed various computer-aided diagnosis (CADx) techniques to tackle liver and tumor segmentation. Recently, 2D and 3D fully convolutional neural network (FCN) based methods~\cite{han2017automatic,li2018h,chlebus2018automatic,vorontsov2018liver,wang2019volumetric} have achieved state-of-the-art performance. 3D FCN models are supposed to consider the contexts on the z-axis, which would lead to better segmentation accuracy than 2D models. Nevertheless, the current 3D segmentation models have an excessive number of parameters with extremely high complexity, which are difficulty to train. Notably, 3D models require large size labeled training data and rich computational resources for optimization. 
Furthermore, many CT scan shows a large variation on the z-axis slice spacing ranges since it consists of anisotropic dimensions.
For example, the slice spacing ranges are from 0.45 mm to 6.0 mm in the LiTS challenge dataset~\cite{bilic2019liver}.
A popular solution is to sample the scans into a fixed spacing (e.g. 1.0 mm) during training 3D or 2D/3D hybrid models. However, for cases with a fine spacing of less than 1.0 mm, some important inter-slice information will be lost. For cases with a coarse spacing of larger than 1.0 mm, extra errors will be introduced by sampling. 
To the best of our knowledge, the existing deep learning based methods \cite{han2017automatic,li2018h,chlebus2018automatic,vorontsov2018liver,wang2019volumetric} did not explicitly model the edges of objects (liver and tumor). The complementarity between the objects and their edges has not been explored, which might boost the segmentation performance.

To address these problems, we propose a powerful 2D segmentation model for liver and tumor segmentation by leveraging and emphasizing their edge information as complementary information. 
We investigate the correlation between liver/tumor segmentation and edge prediction.
The proposed model has three main contributions: 1) Our segmentation model is trained with an edge enhanced cost function, which explicitly models complementary and discriminative feature information within the network to preserve the liver and tumor boundaries.
2) A deep cross feature fusion module is proposed to bidirectionally refine multi-scale features from both objects (\ie liver and tumor) and their edges.
3) Extensive experiments on the publicly available LiTS and 3DIRCADb datasets show the superiority of the proposed method as compared to several state-of-the-art 2D, 3D and hybrid models for liver and tumor segmentation.

\section{Methodology}

\begin{figure*}[t!]
  \centering
  \includegraphics[width=\linewidth]{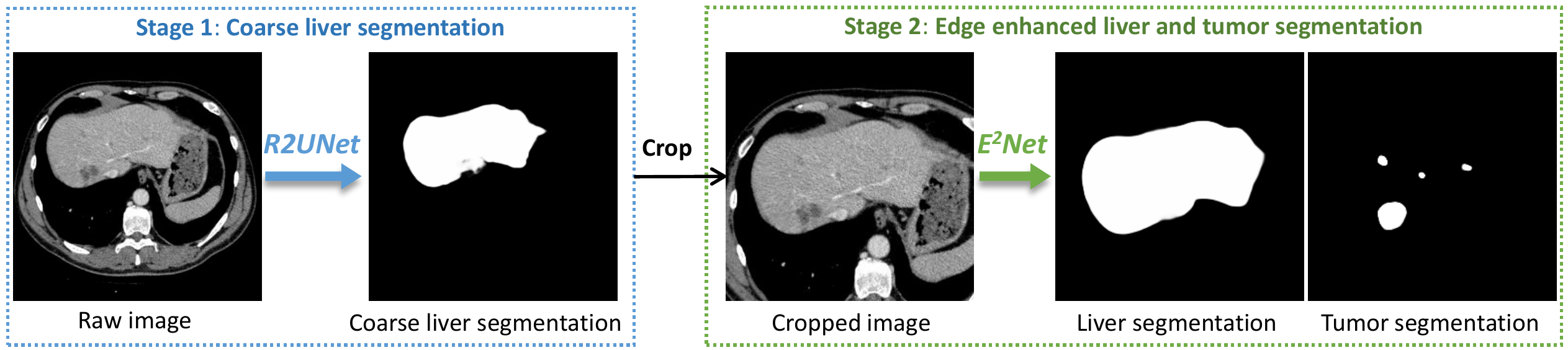}
  \caption{Overview of the proposed framework.}
  \label{fig:framework}
\end{figure*}

Fig. \ref{fig:framework} shows an overview of the proposed framework, which consists of two stages. In the first stage, given a CT image, we coarsely segment the liver region using a fully convolutional neural network (denoted R2UNet) if a liver appears. Based on the segmented region, a CT sub-image is cropped as the input of the second stage, wherein we propose an edge enhanced network (E$^2$Net) that is used to simultaneously segment the liver and tumor regions more accurately.

\begin{figure*}[t!]
  \centering
  \includegraphics[width=\linewidth]{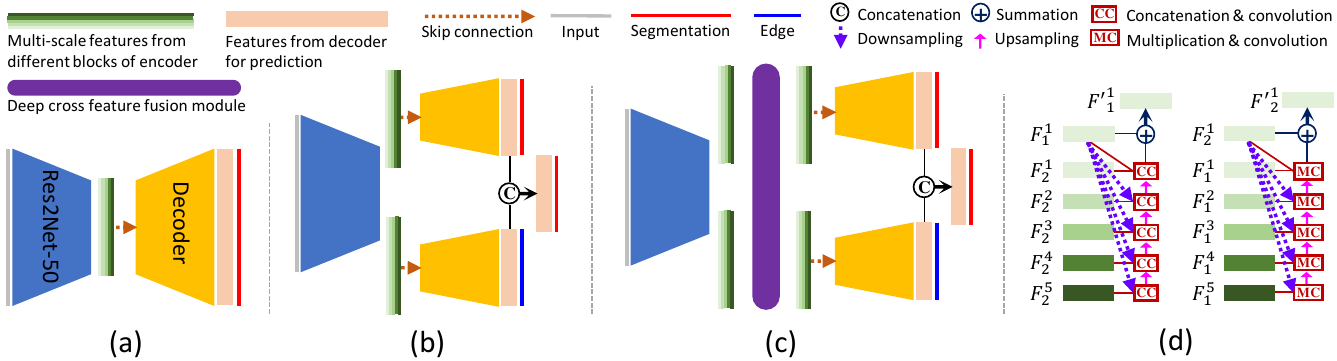}
  \caption{Illustration of different architecture configurations: (a) R2UNet with a single branch used as a baseline, using liver mask as supervision; (b) Adding an extra branch using edge as supervision, performing feature fusion with concatenation; (c) Adding the proposed deep cross feature fusion (DCFF) module for feature refinement, \ie the proposed E$^2$Net; (d) Illustration of a single scale feature refinement using DCFF.}
  \label{fig:network}
\end{figure*}

\textbf{Coarse Liver Segmentation.}
We use Res2Net-50 \cite{gao2019res2net} as the backbone to extract multi-scale features from CT images.
Res2Net \cite{gao2019res2net} can represent multi-scale features at a granular level and increase the range of receptive fields for each network layer. It has been demonstrated that Res2Net blocks brought consistent performance gains over baseline models, \eg ResNet \cite{he2016deep}, for various tasks such as semantic segmentation.
For an input CT image with a size of $H \times W$, the multi-scale features extracted using Res2Net-50 are denoted as $F=\{F^i|i=1,2,3,4,5\}$. The size of $F^i$ is $ \frac{H}{2^i} \times \frac{W}{2^i} \times C_i $, where $C_i$ is the number of channels. To reduce the computational cost, we pad two convolutional layers with 32 $1\times1$ and $3\times3$ kernels to each feature $F^i$ to compress it to 32 channels. All compressed features are sent to a decoder similar to UNet \cite{ronneberger2015u} except all convolutional layers having 32 kernels for coarse liver segmentation. We denote this network as R2UNet. Please refer to Fig. \ref{fig:network}(a) for its architecture illustration.

\textbf{Accurate Liver and Tumor Segmentation Using E$^2$Net.}
Based on the coarse liver segmentation result, a sub-image is cropped, where most of the irrelevant regions are removed. Using the sub-image as input, the network used in the second stage can focus on learning discriminative features for accurate liver and tumor segmentation.
A baseline method is to use R2UNet with two output channels, one for liver and the other for tumor segmentation.
We empirically find that this baseline cannot well segment the areas near the boundaries of the liver and tumor when the boundaries are blurry. To alleviate this issue, we introduce an extra branch to explicitly learn features for edge prediction. This branch has the same architecture as the aforementioned R2UNet segmentation model. We observe that the features extracted by these two branches retain complementary information. A straightforward strategy is to fuse these features by concatenating them channel-wisely for the final liver and tumor segmentation. Please refer to Fig. \ref{fig:network}(b) for the illustration of this improved architecture.

Using edge as supervision, the heavy imbalance between edge and other pixels hinders the model from learning highly discriminative features for high-quality edge prediction. 
A weighted loss can be used to alleviate this issue. But we provide a new solution from a totally different perspective. We first perform distance transformation on the edge image to get a distance map. Then we multiply it with the binary liver or tumor mask and normalize it to [0, 1]. Finally, the result of 1 minus the normalized distance map is used as supervision, where the pixels closer to the edge have larger values. The intuition behind is that the pixels closer to the edge are more difficult to segment.

As shown in Fig. \ref{fig:network}(b), there is no interaction between the features from two branches except the final fusion step. However, there should be some interrelations between the compressed multi-scale features extracted by the segmentation branch (denoted $F_1=\{F_1^i|i=1,2,3,4,5\}$) and that extracted by the edge prediction branch (denoted $F_2=\{F_2^i|i=1,2,3,4,5\}$). For example, compared with $F_2^i$,  $F_1^i * F_2^i$ could better represent the edge information by suppressing the features from non-edge areas. Compared with $F_1^i$, $F_1^i + F_2^i$ could better represent the liver or tumor region by suppressing the features from the background. To introduce these interrelations, we propose a deep cross feature fusion (DCFF) module to refine the multi-scale features $F_1$ and $F_2$. 
Our final model E$^2$Net for accurate liver and tumor segmentation in the second stage is shown in Fig. \ref{fig:network}(c).

In the DCFF module, for feature of each scale from one branch, we use all equal and larger scale features from the other branch to refine it. Let's take $F_1^1$ and $F_2^1$ as examples. For $F_1^1$, its refined feature ${F'}_1^1$ is obtained based on $F_1^1$ itself and all features $F_2=\{F_2^i|i=1,2,3,4,5\}$. As shown in Fig. \ref{fig:network}(d), for each $F_2^i$, we first downsample $F_1^1$ to the same size as $F_2^i$ and concatenate them. Then a series of operations including concatenation, convolution and upsampling are performed like the behavior of R2UNet's decoder. Finally, the above resulted feature and $F_1^1$ are summed together to obtain ${F'}_1^1$. For $F_2^1\Rightarrow{F'}_2^1$, it is similar to $F_1^1\Rightarrow{F'}_1^1$ except replacing all concatenation operations with element-wise multiplication. The discriminability of the multi-scale features is supposed to be improved by this inter-relational refinement.

 Note that although the encoders and decoders in both stages have the same structures, their weights are not shared so that each model will avoid learning highly correlated information. This is also validated by experiments.

\textbf{Model Optimization.}
For both segmentation and edge prediction tasks, we use binary cross entropy (BCE). It is defined as:
\begin{equation}
    \ell_{BCE}(g,p)=-\sum\nolimits_{(x, y)}[g_{x, y} \log (p_{x, y})+(1-g_{x, y}) \log (1-p_{x, y})]
\end{equation}
where $p_{x, y}$ and $g_{x, y}$ are prediction and ground truth of the pixel $(x, y)$. As a pixel-wise loss, BCE does not consider the global structure of the object. To deal with this, the IoU loss \cite{iouloss} aims to optimize the global structure of the segmented object rather than focusing on a single pixel. It is defined as:
\begin{equation}
    \ell_{IoU}(g,p)=1-\frac{\sum_{(x, y)}[g_{x, y} * p_{x, y}]}{\sum_{(x, y)}[g_{x, y}+p_{x, y}-g_{x, y} * p_{x, y}]}
\end{equation}

Hence, the objective of the first stage is defined as the summation of the IoU loss and the BCE loss. It is formulated as $\ell_{1^{st}}=\ell_{IoU}(g,s)+\ell_{BCE}(g,s)$,
where $g/s$ is the liver ground truth mask/segmentation result. The second stage has three outputs, \ie $s^1$, $e$ and $s^2$, from the segmentation branch, the edge prediction branch, and the fusion step. The objective of this stage is defined as:
\begin{equation}
    \ell_{2^{nd}}=\ell_{IoU}(g_m,s^1)+\ell_{BCE}(g_m,s^1)+\ell_{IoU}(g_m,s^2)+\ell_{BCE}(g_m,s^2)+4*\ell_{BCE}(g_e,e)
\end{equation}
where $g_m$/$g_e$ is the liver and tumor ground truth masks/edge distance maps.

The models in the two stages are trained separately. We use stochastic gradient descent with a momentum of 0.9, an initial learning rate of 0.01, which is divided by 10 once the validation loss is stable. After decreasing the learning rate twice, we stop training. The training batch size is 32. For the first stage, we randomly cropped $448 \times 448$ sub-images from the whole CT images during training. For the second stage, we randomly pad the liver regions with 10 to 60 pixels and resize them into $256 \times 256$. During inference, we segment all 2D slices of the input CT scan and stack the results to get the 3D segmentation. The whole training process costs about 80 hours with an Intel Xeon Gold 6230 CPU and a Tesla V100 GPU. The average inference time is about 12 seconds per case on the 3DIRCADb dataset \cite{soler20103d} when the test batch size is 1.

\section{Experimental Results}

\textbf{Datasets.}
We tested the proposed method on the MICCAI 2017 Liver Tumor Segmentation Challenge~\cite{bilic2019liver} (LiTS dataset) and 3DIRCADb dataset~\cite{soler20103d}. The LiTS dataset contains 131 and 70 contrast-enhanced 3D abdominal CT scans for training and testing, respectively. 
The pixel-wise liver and tumor ground truths are publicly available for the training set while they are withheld for the test set for online evaluation. The 3DIRCADb dataset contains 20 venous phase enhanced CT scans, where 15 scans have hepatic tumors in the liver. The range of the original Hounsfield Unit (HU) values in these CT scans is from less than -3,000 to more than +3,000. We truncated the image intensity values of all scans to the range of [-100, 240] HU to remove the irrelevant details, and then normalized them to [-1, 1]. The 131 LiTS training samples were randomly split into a training set (111) and a validation set (20) for model training and selection. The 3DIRCADb dataset was used for independent hold-out evaluation.

\textbf{Evaluation Criteria.} The output of the fusion step ($s^2$) is used as the final liver and tumor segmentation result for performance evaluation. Following the evaluation of the LiTS challenge and other approaches~\cite{chlebus2017neural,li2018h,chlebus2018automatic}, we used the Dice per case score and the global Dice score (Dice global) to evaluate the segmentation performance, and also used the root mean square error (RMSE) to measure the tumor burden that is defined as the liver/tumor ratio \cite{tumorburden}. Dice per case score refers to an average Dice score per CT scan/volume while global Dice score is the Dice score evaluated by combining all CT scans into one.

\begin{figure*}[t!]
	\begin{minipage}[b]{1.0\linewidth}
		\centering
		\includegraphics[width=0.99\linewidth]{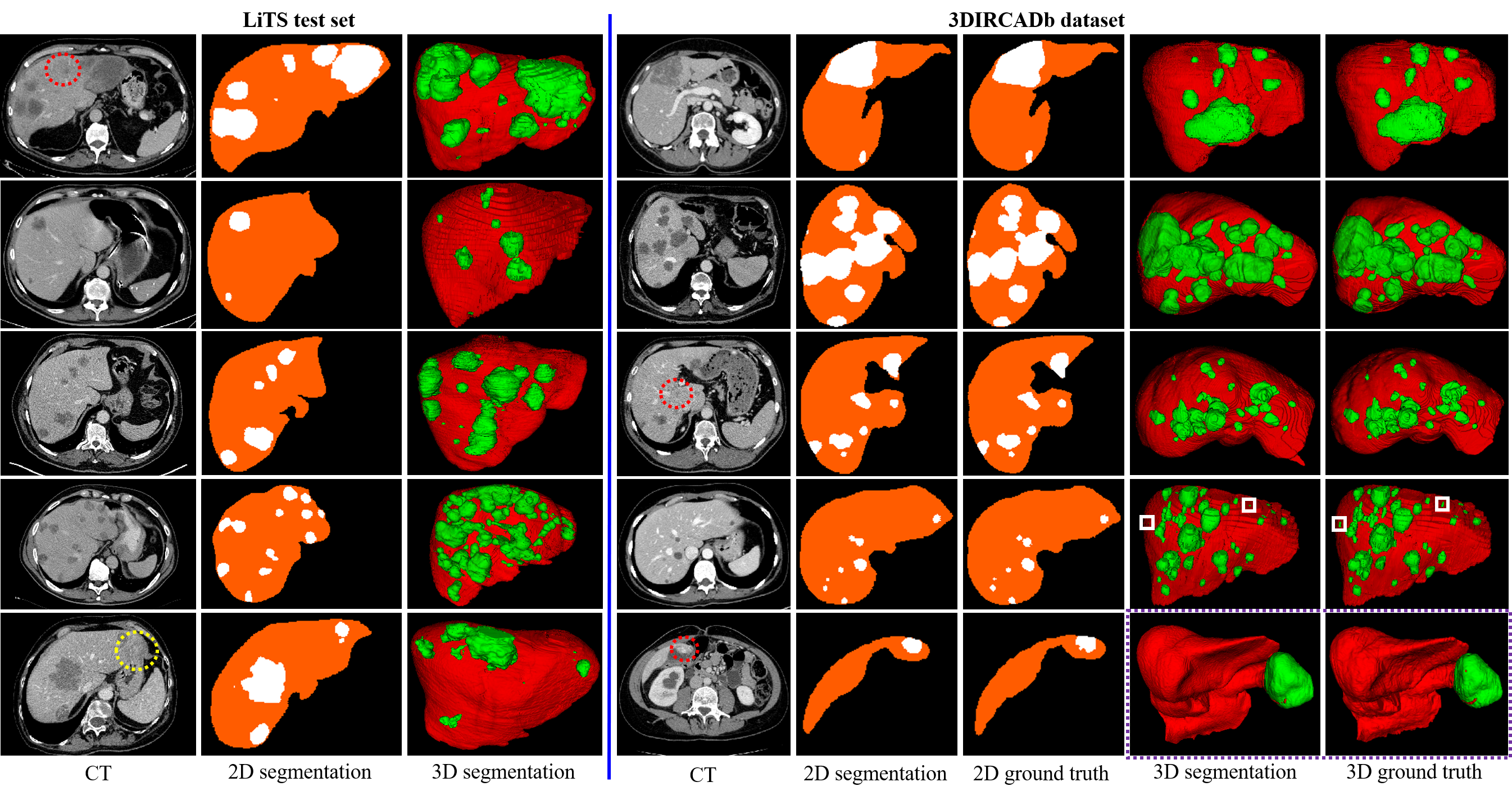} \\
	\end{minipage}
    \begin{minipage}[b]{0.375\linewidth}
        \label{fig:result-a}
		\centering
		\centerline{(a)}\medskip
	\end{minipage}
	\begin{minipage}[b]{0.615\linewidth}
	    \label{fig:result-b}
		\centering
		\centerline{(b)}\medskip
	\end{minipage}
	\caption{Visual examples of segmentation results produced by the proposed method on (a) LiTS test set and (b) 3DIRCADb dataset. 
	The ground truths are unavailable in the LiTS test set. To make the tumors visible, we set a larger transparency to livers in all 3D examples except the ones indicated by a purple box where the livers' 3D shapes are well visualized. The yellow/red dashed circles indicate some cases where the livers/tumors' boundaries are blurry. The white boxes in last two columns indicate two cases where the small tumors cannot be well segmented. Best viewed in color.}
	\label{fig:result}
\end{figure*}

\textbf{Qualitative Segmentation Results.}
\label{example}
Fig. \ref{fig:result} shows five visual examples of liver and tumor segmentation results produced by the proposed E$^2$Net on both LiTS test set and 3DIRCADb dataset. From Fig. \ref{fig:result}, the liver and tumor regions can be well segmented by our E$^2$Net, even for some cases where the boundaries of the livers (the yellow dashed circle) or tumors (the red dashed circles) are very blurry. 
This is an advantage of E$^2$Net that models edge information within the network training to preserve liver and tumor boundaries.
In Fig. \ref{fig:result}(b), the automatic segmentation results are very close to the ground truths. From the example indicated by a purple box, the 3D surface of our segmentation is smoother than the ground truth, suggesting that E$^2$Net is able to correct some inaccuracies of the human annotations. Therefore, the annotators can simply refine the automatic segmentations to get high-quality annotations. As such, E$^2$Net has the potential to speed up the manual annotation process. In the last two columns of Fig. \ref{fig:result}, some small tumors indicated by white boxes cannot be well segmented. Through investigation, we find that these tumors are small and have low contrast with their surrounding liver regions. For such cases, providing only the 2D in-plane information might be insufficient for E$^2$Net to learn discriminative features to distinguish them. One possible solution is to incorporate 3D spatial and contextual information between sequential slices, which is left as future work.

\begin{table}[t!]
	\begin{center}
		\caption{Segmentation results produced by different methods on the LiTS test set.}
		\label{tab:lits}
		{
			\scriptsize
			\begin{tabu} to 0.995\textwidth {| X[0.4,c] | X[2.2,c] | X[0.7,c] | X[0.7,c] | X[0.7,c] | X[0.7,c] | X[1.1,c] |}
				\hline
				\multicolumn{1}{|c}{}& & \multicolumn{2}{c|}{Liver Dice (\%)} & \multicolumn{2}{c|}{Tumor Dice (\%)} & Tumor Burden \\ \cline{3-7}
                 \multicolumn{2}{|c|}{\multirow{-2}{*}{Methods}} & Per case & Global & Per case & Global & RMSE \\ \hline \hline
                & TwoFCNs \cite{vorontsov2018liver} & 95.1 & 95.1 & 66.1 & 78.3 & 0.023 \\
                & DeeplabV3+ \cite{chen2018encoder} & 95.7 & 96.1 & 66.6 & 80.4 & 0.016 \\
                & 2.5DResUNet \cite{han2017automatic} & - & - & 67.0 & - & - \\
                & UNet+SP \cite{chlebus2018automatic} & 96.0 & 96.5 & 67.6 & 79.6 & 0.020 \\
                & DenseNet (pre-trained) \cite{li2018h} & 95.3 & 95.9 & 68.3 & 81.8 & - \\
                & DenseUNet (pre-trained) \cite{li2018h} & 95.8 & 96.3 & 70.2 & 82.1 & - \\
                \multirow{-5}{*}{2D} & E$^2$Net (Ours) & 96.4 & \textbf{96.8} & 72.4 & \textbf{82.9} & \textbf{0.015} \\
                \hline \hline
                & UNet \cite{chlebus2017neural} & - & - & 65.0 & - & - \\
                & H-DenseUNet \cite{li2018h} & 96.1 & 96.5 & 72.2 & 82.4 & \textbf{0.015} \\
                \multirow{-3}{*}{2\&3D} & LW-HCN \cite{zhang2019light} & \textbf{96.5} & \textbf{96.8} & \textbf{73.0} & 82.0 & \textbf{0.015} \\
                \hline \hline
                & DenseUNet \cite{li2018h} & 93.6 & 92.9 & 59.4 & 78.8 & - \\
                & I3D \cite{carreira2017quo} & 95.7 & 96.0 & 62.4 & 77.6 & 0.025 \\
                \multirow{-3}{*}{3D} & I3D (pre-trained) \cite{carreira2017quo} & 95.6 & 96.2 & 66.6 & 79.9 & 0.023 \\
				\hline
			\end{tabu}
		}
	\end{center}
\end{table}

\textbf{Quantitative Segmentation Results.}
Table \ref{tab:lits} lists the liver and tumor segmentation results of the proposed E$^2$Net and other state-of-the-art methods published in the existing literature on the LiTS test set. As a 2D model, our method is first compared with the other 2D models. We can see from the table that it outperforms the others by a large margin. For instance, the tumor Dice per case score is improved from 70.2\% to 72.4\%, suggesting that our model is capable of learning more discriminative features from 2D CT slices for better segmentation.
Although 3D models can consider the 3D spatial information from the CT scans, they have much more trainable parameters than 2D models that require a lot more training data for optimization. However, the amount of annotated training data is insufficient in practice for this task. Therefore, it is not surprising that the 3D models even perform worse than the existing 2D models, as shown in Table \ref{tab:lits}. To address this problem, researchers have developed some hybrid models to consider both 2D in-plane and 3D spatial information, which achieved better performance than a single 2D or 3D model, such as H-DenseUNet \cite{li2018h} and LW-HCN \cite{zhang2019light}. As shown in Table \ref{tab:lits}, our E$^2$Net is better than H-DenseUNet in terms of all evaluation metrics and gets comparable performance with LW-HCN. This demonstrates that a powerful and well-designed 2D model, such as the proposed method, still can compete with the hybrid models, even without using 3D information. For tumor segmentation, the global Dice score of our method is better than the one of LW-HCN, but the Dice per case score is worse. One possible reason we find is that there are some CT scans only having small tumors while our method cannot well segment small tumors sometimes as described in Section \ref{example}. For such CT scans, our method gets low Dice scores that heavily reduce the average Dice score per case. However, these small tumor segmentations only make a small contribution to the global Dice score computation.

Table \ref{tab:3dircadb} lists the liver and tumor segmentation results of different methods in terms of Dice per case score on the 3DIRCADb dataset. Our method is trained using the LiTS training set and directly tested on the 3DIRCADb dataset, meaning that the CT scans of 3DIRCADb dataset are totally unseen by our E$^2$Net and different from the LiTS dataset. From Table \ref{tab:3dircadb}, we can see that E$^2$Net obtains the best performance on both liver and tumor segmentation. Compared with the hybrid model LW-HCN \cite{zhang2019light} that has the highest tumor Dice per case score of 94.1\% in the literature, our 2D model improves this score to 95.7\% with a large margin. These remarkable results demonstrate that our method has strong generalizability to accurately segment the livers and their tumors.

\begin{table}[!t]
  \begin{center}
    {
		\caption{Results produced by different methods in terms of Dice per case (\%) on the 3DIRCADb dataset.}
		\label{tab:3dircadb}
        \begin{tabu} to 0.7\textwidth {| X[2.5,c] | X[c] | X[c] |}
          \hline
          Methods & Liver & Tumor \\ \hline
          \hline
          Moghbel et al \cite{moghbel2016automatic} & 91.1 & 75.0 \\
          Foruzan et al \cite{foruzan2016improved} & - & 82.0 \\
          Wu et al \cite{wu20173d} & - & 83.0 \\
    	  H-DenseUNet \cite{li2018h} & 98.2 & 93.7 \\
    	  LW-HCN \cite{zhang2019light} & 98.1 & 94.1 \\
    	  E$^2$Net (Ours) & \textbf{98.9} & \textbf{95.7} \\
          \hline
          \end{tabu}
    }
    \end{center}
\end{table}%

\begin{table}[!t]
    \begin{center}
    {
		\caption{Results of different configurations of our method in terms of Dice per case (\%) on the LiTS validation set.}
		\label{tab:ablation}
        \begin{tabu} to 0.7\textwidth {| X[2.5] | X[c] | X[c] |}
              \hline
              \centering Configurations & Liver & Tumor \\ \hline
              \hline
              \centering 1$^{st}$ stage & 95.3 & -\\ \hline
              \centering Baseline (2$^{nd}$ stage) & 96.1 & 71.4 \\ 
        	  \MyIndent\MyIndent\MyIndent\MyIndent + edge & 96.6 & 72.2 \\
        	  \MyIndent\MyIndent\MyIndent\MyIndent + dist & 97.1 & 72.9 \\
        	  \MyIndent\MyIndent\MyIndent\MyIndent + edge + DCFF & 97.5 & 73.7 \\
        	  \MyIndent\MyIndent\MyIndent\MyIndent + dist \ + DCFF & \textbf{97.8} & \textbf{74.8} \\
              \hline
        \end{tabu}
    }
    \end{center}
\end{table}%

\textbf{Ablation Study.}
To explore the effect of different design components in the proposed method, we set up the following different experimental configurations and test them on the LiTS validation set: 1) only using the first stage for liver segmentation, 2) using the model (R2UNet) of the first stage as the baseline in the second stage, 3) adding another branch and using edge or distance map (dist) as extra supervision, 4) using the proposed deep cross feature fusion (DCFF) module. Table \ref{tab:ablation} lists the liver and tumor segmentation performance of different configurations in terms of the Dice per case score. From this table, we can see that (1) the second stage is essential to get better performance, meaning that such a coarse-to-fine strategy is effective for liver and tumor segmentation. (2) \textit{Baseline + edge/dist} is better than \textit{Baseline}, suggesting that the added branch can learn complementary features to the baseline for performance improvement. (3) As extra supervision, the distance map performs better than the edge, demonstrating the effectiveness of our designed distance map. (4) The largest performance boost is obtained when using DCFF, suggesting that the proposed DCFF can fuse and refine the multi-scale features from both branches effectively and simultaneously. We observed the same findings as above when using E$^2$Net in both stages. Hence, in our experiment, we used a simpler R2UNet model in the first stage.

\section{Conclusions}
This paper proposes an edge enhanced deep learning network for robust and accurate liver/tumor segmentation. Operating on original 2D CT slices, the proposed method eliminates the z-axis re-sampling errors caused by different CT slice thickness in other 3D segmentation models. Moreover, our model is easier to train/apply with less computational resources compared to 3D models.
By enhancing the edge information, the proposed method improves the performance of liver and tumor segmentation significantly, especially when poor boundaries exist in CT images.
Extensive experiments on the challenging LiTS and 3DIRCADb datasets demonstrate the power and effectiveness of our method.
Future work will explore the possibility of developing low cost and fast segmentation method which can be used in an embedded system on CT scanner. 

\noindent\textbf{Acknowledgments.}
This research was supported by the Intramural Research Program of the National Institutes of Health Clinical Center and by the Ping An Insurance Company through a Cooperative Research and Development Agreement. We thank Nvidia for GPU card donation.

%
%
%
\bibliographystyle{ieeetr}
\bibliography{ref}
%
%
\end{document}